\documentclass[english,aps,prl,twocolumn,superscriptaddress,showpacs, amsmath]{revtex4-2}
\usepackage{amsmath}
\usepackage[T1]{fontenc}
\usepackage{graphicx}
\usepackage{amssymb}
\usepackage{dcolumn}
\usepackage{color}
\usepackage{bm}
\usepackage[normalem]{ulem}
\usepackage{url}
\usepackage[ colorlinks = true,
            linkcolor = red,
            urlcolor  = red,
            citecolor = blue,
            anchorcolor = red]{hyperref}
\usepackage[colorinlistoftodos]{todonotes}
\usepackage{bibentry}

\usepackage{lineno}
\usepackage{pdfpages}
\makeatletter
\AtBeginDocument{\let\LS@rot\@undefined}
\makeatother

\begin{document}

\title{Electrically switchable tunneling across a graphene pn junction: evidence for canted antiferromagnetic phase in $\nu=0$ state
}

\author{Arup Kumar Paul$^1$, Manas Ranjan Sahu$^1$, Kenji Watanabe$^{2}$, Takashi Taniguchi$^{2}$, J. K. Jain$^{3}$, Ganpathy Murthy$^{4, \star}$, and Anindya Das$^{1, \dagger}$}

\affiliation{Department of Physics,Indian Institute of Science, Bangalore, 560012, India.}
\affiliation{National Institute for Materials Science, Namiki 1-1, Ibaraki 305-0044, Japan.}
\affiliation{Department of Physics, The Pennsylvania State University, University Park, Pennsylvania 16802, USA.}
\affiliation{Department of Physics and Astronomy, University of Kentucky, Lexington, Kentucky 40506, USA.}


\begin{abstract}
The ground state of a graphene sheet at charge neutrality in a perpendicular magnetic field remains enigmatic, with various experiments supporting canted antiferromagnetic, bond ordered, and even charge density wave phases. A promising avenue to elucidating the nature of this state is to sandwich it between regions of different filling factors, and study spin-dependent tunneling across the edge modes at the interfaces. Here we report on tunnel transport through a $\nu=0$ region in a graphite-gated, hexagonal boron nitride ($hBN$) encapsulated  monolayer graphene device, with the $\nu=0$ strip sandwiched by  spin-polarized $\nu=\pm1$ quantum Hall states. We observe finite tunneling ($t \sim 0.3-0.6$) between the $\nu=\pm1$ edges at not too small magnetic fields ($B>3T$) and low tunnel bias voltage ($<30-60\mu V$), which is surprising because electrons at the edge states nominally have opposite spins. 
Hartree-Fock calculations elucidate these phenomena as being driven by the formation of a CAF order parameter in the $\nu=0$ region at zero bias (for wide enough junctions) leading to non-orthogonal spins at the edges. Remarkably, this tunneling can be controllably switched off by increasing bias; bias voltage leads to a pileup of charge at the junction,  leading to a collapse of the CAF order and a suppression of the tunneling.

\end{abstract}

\maketitle

\begin{figure}[ht!]
\includegraphics[width=0.42\textwidth]{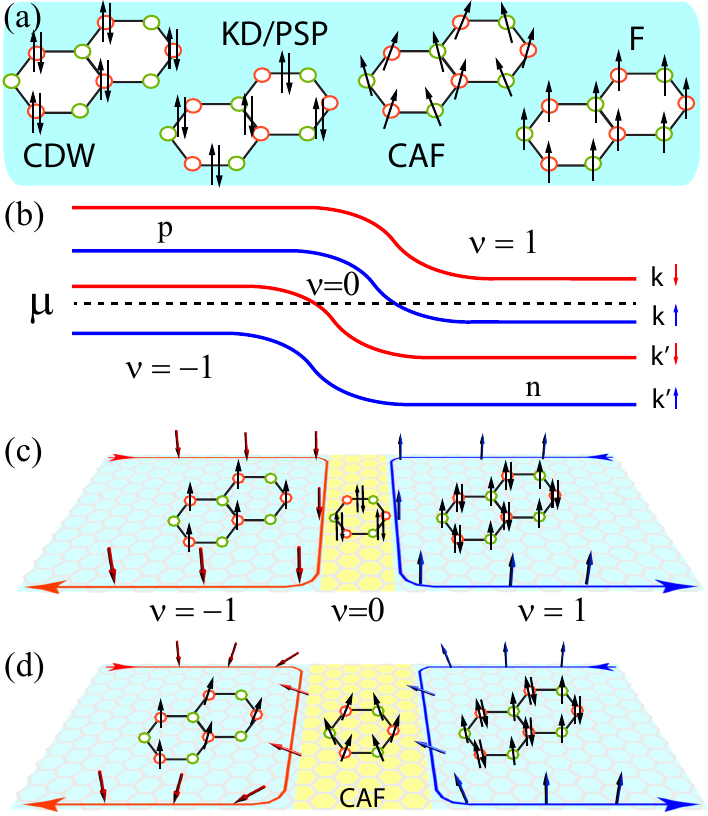}
\caption{\label{fig1} \textbf{(a)}  Ground states of the half filled ($\nu=0$) ZLL, with the distribution and spin ($\uparrow$/$\downarrow$-arrows) of the electrons between the two sublattices (red and green circles). From left to right: charge density wave (CDW), Kekule  Distorted (KD) or partially sublattice polarized phase (PSP),  canted anti-ferromagnet (CAF)  and ferromagnet (F). \textbf{(b)} Symmetry broken LLs 
across a graphene pn junction in the CDW/KD state. The colors corresponds to the nominal spin polarization and $K$ or $K'$ represents the valley index of the LLs. \textbf{(c)} and \textbf{(d)} Schematic of a graphene pn junction  without/with the CAF ordering of the $\nu=0$ region (yellow), respectively. The red (blue) arrows show spin polarization of the hole (electron)-like edge states. The black arrows show sublattice spin polarization inside the bulk.}
\end{figure}

The particle-hole symmetric band-structure of monolayer graphene with spin, valley or sublattice symmetries has been proposed to give rise to a rich variety of interaction-driven symmetry-broken quantum Hall (QH) phases~\cite{yang2010hierarchy,abanin2006spin,weitz2010broken,zhao2010symmetry,feldman2009broken,maher2014tunable,papic2014topological,zibrov2017tunable,tHoke2006fractional,apalkov2006fractional,bolotin2009observation,dean2011multicomponent,du2009fractional,feldman2012unconventional,kou2014electron}. 
The simplest  symmetry-broken states in the four-fold (nearly) degenerate manifold of the $n=0$ Landau levels (the zero-energy Landau levels, or ZLLs) are described within the formalism of quantum Hall ferromagnetism (QHFM)~\cite{Sondhi1993,Yangetal1994,Moonetal,Yangetal2006,nomura2006quantum}. In the ZLLs, a plethora of phases have been observed~\cite{nomura2006quantum,young2012spin,yu2014hierarchy,kim2021edge,alicea2006graphene,young2014tunable,goerbig2011electronic}, but the most enigmatic QH phase arises at the charge neutrality point ($\nu=0$)~\cite{sarma2009enigma,Herbut1,Herbut2,knothe2015edge,kharitonov2012canted,gorbar2008dynamics,kharitonov2012edge,kharitonov2012phase,pezzini2015canted,li2020transition,kim2021edge,zibrov2018even,STM_Yazdani2021visualizing,coissard2021imaging}.  While the Coulomb interaction has an $SU(4)$ spin-valley symmetry, residual interactions~\cite{alicea2006graphene} break this symmetry and  determine the physical properties of the $\nu=0$ ground state~\cite{kharitonov2012phase,kharitonov2012canted}, the  possibilities being  charge density wave (CDW),  bond ordered/Kekule distorted (KD),  canted anti-ferromagnet (CAF), and  fully spin-polarized  (F) phases (see Fig.~\ref{fig1}a)~\cite{kharitonov2012phase}. All these phases except F, which is a quantum spin Hall insulator, are  
insulators without protected edge states, making their experimental identification difficult.

Experimentally, a phase transition of the $\nu = 0$ phase from an ordinary insulator to a quantum spin Hall insulator was observed~\cite{young2014tunable} as a function of tilted field (the Zeeman energy $E_Z$), with the high-$E_Z$ phase yielding a two-terminal conductance of $G_{2T}\approx2e^2/h$, consistent with the F phase.  At purely perpendicular field, magnon transmission experiments~\cite{Magnontransmission2018,Young_Skyrmion_Solid_Graphene_2019,zhou2021strong,Assouline_2021} show that spin excitations do traverse a $\nu=0$ region, which must therefore have some form of spin order, suggesting the CAF state. The spin excitations are generated in fully spin-polarized regions~\cite{Magnontransmission2018,Young_Skyrmion_Solid_Graphene_2019,zhou2021strong,Assouline_2021}, necessarily have energies above $E_Z$, and are subject to kinematic constraints~\cite{Huang_MacDonald2021}. However, more recent scanning tunneling microscopy/spectroscopy (STM) experiments find evidence for 
bond order~\cite{li2019:stm,STM_Yazdani2021visualizing,coissard2021imaging}, and also CDW order~\cite{STM_Yazdani2021visualizing,coissard2021imaging}; no information regarding magnetic order is available from the STM experiments. While past theory~\cite{kharitonov2012phase} implies that the $\nu=0$ state is either CAF or bond ordered, recent theoretical work supports the coexistence of  CAF and bond order~\cite{Das_Kaul_Murthy_2022}. Given the tension between different experimental observations, elucidating the order parameters in the $\nu=0$ ground state of graphene is of utmost importance, not least because of the 
possible technological applications of the CAF state.~\cite{yang2006collective,atteia20214,takei2016spin,zhou2021strong,fu2021gapless,Young_Skyrmion_Solid_Graphene_2019,wei2021scattering,wu2020collective,stepanov2018long,wei2018electrical}. We note in this context that the recent observation of a linearly dispersing magnon excitation~\cite{fu2021gapless} in the  $\nu=0$ state in a sister material, namely Bernal-stacked bilayer graphene, confirms the presence of CAF order in that material.

We pursue in this work an alternative 
approach to probe the order of the $\nu = 0$ state, by coupling it to better-understood quantum Hall states. A graphene pn junction forms an ideal probe, because the $\nu=0$ region naturally appears at the boundary separating the electron and hole like QH states ~\cite{zimmermann2017tunable,amet2014selective,wei2017mach} as shown in Fig.~\ref{fig1}. In Fig.~\ref{fig1}b, the sublattice symmetry broken LLs 
originating from the ZLL are shown across a graphene pn junction. There is experimental evidence~\cite{VP_Goldhaber_Gordon_2013,zibrov2018even} as well as theoretical support~\cite{VP_Sheng_2007,VP_Macd_2015} for sublattice symmetry breaking due to the encapsulating $hBN$, which makes these states valley polarized. In this letter, we will assume the $\nu=\pm1$ states to be valley polarized. The colors in Fig.~\ref{fig1}b (blue or red) represent the spin-polarization ($\uparrow$/$\downarrow$) of the LLs. The edge states of the electron-like and hole-like states appear at the LL and chemical potential ($\mu$) crossings. In  Figs.~\ref{fig1}c and ~\ref{fig1}d, the $\nu=0$ region is the yellow strip separating the electron-like and hole-like QH states.
The $\nu=\pm1$ states are known to be fully spin polarized in the bulk as shown by the black arrows in Fig.~\ref{fig1}c. 

The essential physics motivating this work is as follows. If the $\nu=0$ region is in a singlet insulator, such as the KD or the CDW phase shown in Fig.~\ref{fig1}c, the $0|1$ edge (blue) has $\uparrow$-spin, while the $-1|0$ edge (red) has $\downarrow$-spin. In this case, one would expect disorder-induced single-particle tunneling between the edges to be heavily suppressed. 
The situation is very different if the $\nu=0$ strip is in a CAF state, as in Fig.~\ref{fig1}d. As is well-known~\cite{takei2016spin,wei2018electrical}, the CAF order can penetrate several magnetic lengths into the bulk of the $\nu=\pm1$ regions (shown by the tilted black arrows in Fig.~\ref{fig1}d), implying that the chiral edge modes of the $\nu=\pm1$ states are no longer fully spin-polarized ( red or blue arrows in Fig.~\ref{fig1}d). In this case, disorder can 
induce tunneling between edges at arbitrarily low bias. For a sufficiently long strip, one can expect roughly half of the input current to tunnel across the strip, 
which would be clear evidence for the existence of  CAF order at $\nu = 0$.

Motivated by these considerations, we have carried out bias-dependent tunneling measurements between the $\nu=\pm1$ edge states co-propagating along a graphene pn junction, and obtained the dependence of the tunneling current on magnetic field. The key finding is the observation of finite tunneling ($t \sim 0.3-0.6$) between $\nu=\pm1$ edge states at and around zero bias at magnetic fields above $3\ T$. We take this to be an evidence for the CAF phase in the $\nu=0$ strip. The absence of tunneling at low perpendicular magnetic fields ($<3T$) is explained, qualitatively, using model Hartree-Fock calculations, which show that at weak fields, the $\nu=0$ strip is too narrow in units of magnetic length $\ell$ to sustain CAF order. We further find an abrupt suppression of tunneling above a critical bias $V_b^*$, which increases with magnetic field ($V_b^*\sim \pm60\mu V$ at 8T). This effect is understood by noting that as the  bias voltage increases, charge piles up at the edges of the $\nu=0$ strip, effectively narrowing it, and eventually destabilizing the CAF state. This allows for switching on and off of the tunneling electrically.

\begin{figure}[ht!]
\includegraphics[width=0.48\textwidth]{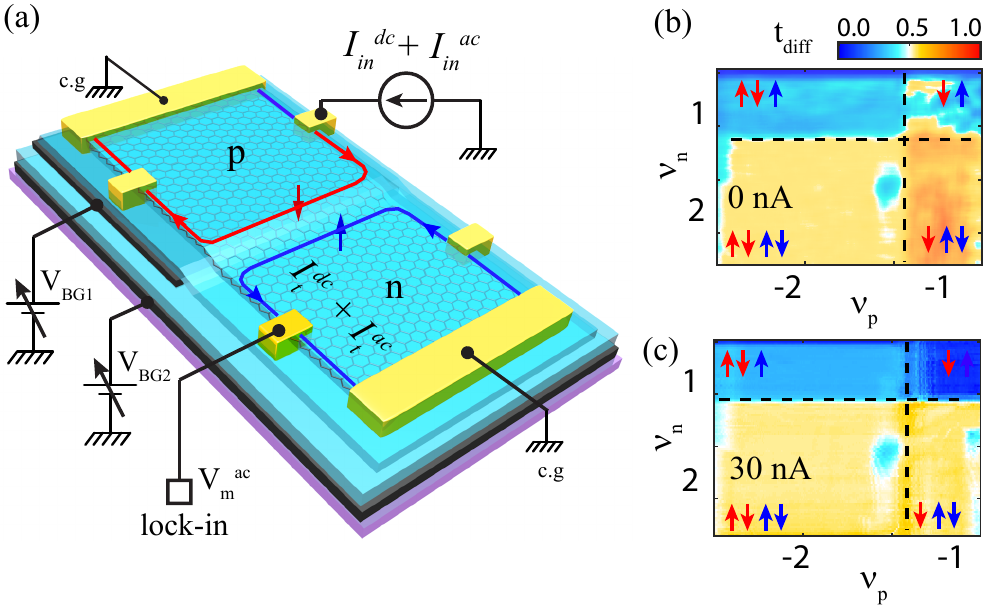}
\caption{\label{fig2} \textbf{Device schematic and junction response} \textbf{(a)} Schematic of the dual gated graphene device with measurement setup. \textbf{(b)} and \textbf{(c)} shows 
$t_{diff}$ as function of the $\nu_p$ and $\nu_n$ at $8\;T$ and $40\;mK$ for $I_{in}^{dc}$=0 nA and $I_{in}^{dc}$=30 nA, respectively. The black dashed lines marks the boundary of the different filling factor plateaus. The spin polarization ($\uparrow$ or $\downarrow$-arrows) of the electron (hole) like edge states on each plateau are shown by the blue (red) arrows. }
\end{figure}

The device schematic and  measurement setup are shown in Fig.~\ref{fig2}(a). The device consists of an $hBN$ encapsulated graphene flake, with two bottom graphite gates BG1 and BG2.  Details about the device structure can be found in our previous work~\cite{paul2020interplay} and the supplementary information  1 (SI-1).
 The measurement-setup is described in SI-2. We have used a six contact Hall bar geometry for the transport measurement, with three contacts on either side of the pn junction. The extreme two contacts are connected to the dilution mixing chamber (MC) plate to serve as cold ground (c.g), and the Hall bars are used for current injection and voltage measurements. To study the inter-edge  tunneling at the junction we have measured the bias-dependent differential transmittance ($t_{diff}$) of the junction. For this, a dc-excitation current $I_{in}^{dc}$ together with a small ac-current $I_{in}^{ac}$ ($\sim 1$ nA at 13 Hz) is injected at the p side Hall bar contact. The corresponding ac component of the transmitted current $I_t^{ac}$ is determined from the voltage drop $V_{m}^{ac}$, measured at the n side as shown in Fig.~\ref{fig2}(a). Here, $V_{m}^{ac} = \frac{h}{|\nu_n| e^2}I_t^{ac}$, where $\nu_n$ is the n side filling factor.  The ratio  $I_t^{ac}/I_{in}^{ac}$ gives the measured $t_{diff}$. The net transmittance ($t=I_t^{dc}/I_{in}^{dc}$) of the junction can be calculated from the $t_{diff}$ vs. $I_{in}^{dc}$ plots and the process is  described in details in SI-3. Most of the measurements have been performed with magnetic fields between $2\;T$-$8\;T$, and at temperature of $40\; mK$.

\begin{figure}[h]
\includegraphics[width=0.45\textwidth]{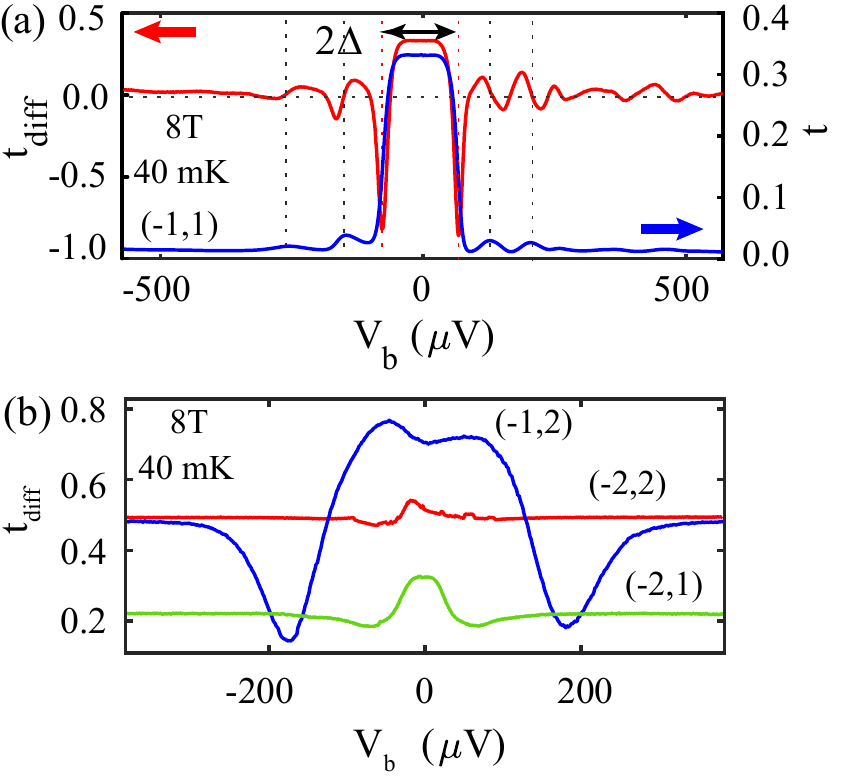}
\caption{\label{fig3}\textbf{Bias dependent transmission:}  \textbf{(a)} $t_{diff}$ (red) and  t (blue) versus bias ($V_b$) for the (-1,1) plateau at $40\; mK$ and $B=8\;T$. The finite $t_{diff}$ or $t$ within a small bias window $2\Delta \; \sim 120\mu V$ (between the red vertical dashed lines) is followed by smaller peaks (vertical black dashed lines) 
and vanishing transmittance at higher bias. \textbf{(b)} $t_{diff}$ vs $V_b$ response for the (-1,2) (blue), (-2,1) (green) and (-2,2) (red) plateau at same temperature and magnetic field.} 
\end{figure}

Figure~\ref{fig2}b shows the measured  $t_{diff}$ (at $B=8$T and $T=40\;mK$) as function of the $p$ and $n$ side filling factors ($\nu_p$ and $\nu_n$, respectively) without dc-excitation current  i.e for $I_{in}^{dc}=0$ nA, while Fig.~\ref{fig2}c shows the $t_{diff}$ for  $I_{in}^{dc}=30$ nA. In both the figures, the black dashed lines demarcate different ($\nu_p,\nu_n$) filling factor plateaus and the ideal spin polarization of the p (n) side edge states on each plateau are shown with the red (blue) arrows.  A comparison of the two figures reveals differences in the $t_{diff}$ magnitudes between the two biasing scenarios at the (-1, 1) and (-1,2) filling factor plateaus. The most striking difference is observed for the ($-1,1$) plateau, where for $I_{in}^{dc}=0$, average $t_{diff}\sim\;0.4$, but for $I_{in}^{dc}=30$ nA, $t_{diff}$ practically vanishes.
 Figure ~\ref{fig3}(a) shows the characteristic bias ($V_{b}$) dependence of $t_{diff}$ (red) or $t$ (blue) for the ($-1,1$) plateau, at $8\;T$ and $40\;mK$. 
  Here, $V_{b}$ corresponds to the net voltage drop across the junction given by $V_{b}=\frac{h}{|\nu_p| e^2}I_{in}^{dc}$. From the plot, three distinct tunneling regimes can be identified: (i) finite tunneling ($t\sim0.4$) within a small bias window $2\Delta$ ($\sim 120\mu V$) around $V_b=0$, as marked by the vertical red dashed lines; (ii)  a sharp fall of $t$ outside the $2\Delta$ window, accompanied by small transmittance peaks shown by the vertical black dashed lines; and (iii) zero transmittance for $V_b>\pm250\mu V$. 
  If the $\nu_{p,n}=\pm1$ edge states have opposite spin polarization as depicted in Fig.~\ref{fig2}b and  \ref{fig2}c, nonzero tunneling is not possible at  $V_b=0$ ~\cite{amet2014selective,wei2018electrical,paul2020interplay}. In this scenario, tunneling between these edge states is only possible when the applied bias exceeds the Zeeman energy required for spin-flip scattering~\cite{wei2018electrical}. Fig.~\ref{fig3}(b) shows the typical $t_{diff}$ vs. $V_b$ responses for other plateaus (also see SI-5). The bias responses at ($-1,2$) and ($-2,1$) plateaus show a low-bias - high-$t$ and a high-bias - low-$t$ region  similar to that observed for the ($-1,1$) plateau. However, for plateaus with $\nu_{p,n}\geq\pm2$ no such bias dependence is observed. In SI-6 we have compared the bias-dependent transmittance for all the plateaus with full or spin-selective equilibration assumed for the QH edge states ~\cite{zimmermann2017tunable,amet2014selective,paul2020interplay}. The comparison shows that while the low bias transmittance matches the full-equilibration model, high bias transmittance agrees with the expected spin-selective equilibration. Thus, the unusual bias response is evidence for  a  state
where the spins of the electron/hole-like  edge-states at the pn junction are not orthogonal at low bias.

\begin{figure}[ht!]\includegraphics[width=0.48\textwidth]{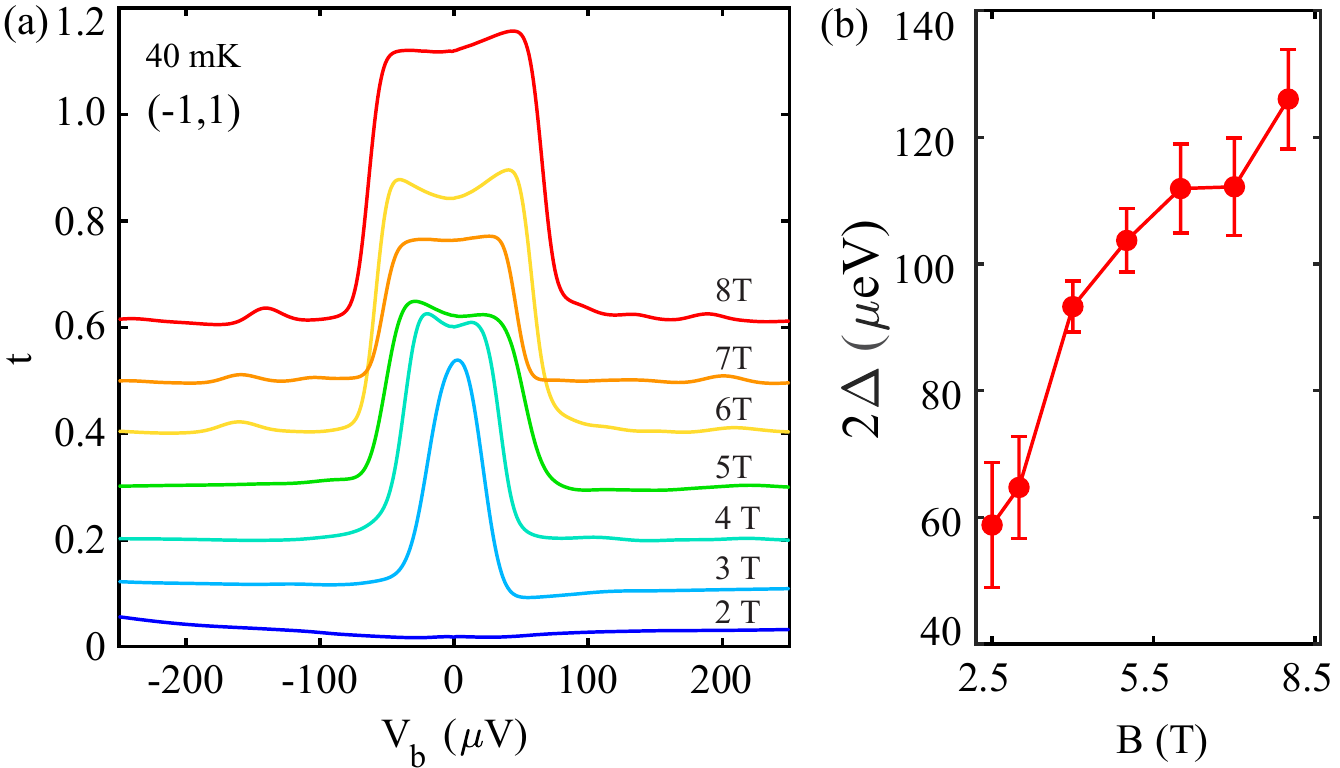}
\caption{\label{fig4}\textbf{Magnetic field  dependence:} \textbf{(a)}  $t$ vs. $V_b$ response as function of magnetic field for the $(-1,1)$ plateau at $40\; mK$.  For clarity, the data has been shifted manually along the $t$ axis with increasing field. \textbf{(b)}  The gap $2\Delta$ (green) for the same plateau as function of B.  The 8T data in (a) are at different gate voltages than the 8T data in the previous figure.
}
\end{figure}

The magnetic field dependence further elucidates the nature of the state  revealed by the bias response. Figure~\ref{fig4}a shows the evolution of the $t$ vs. $V_b$ responses for the ($-1,1$) plateau, with increasing magnetic field and at 40 mK. Here, the bias responses have been vertically offset for clarity. As seen, t is vanishingly small and bias independent at the lowest magnetic-field of $2\;T$. We note that the $\nu=1$ QH plateau is visible for $B=1\;T$, while the $\nu=-1$ QH plateau appears at $2\;T$. At $2\;T$ the absence of tunneling can be attributed to orthogonal spin polarization of the $\nu_{p,n}=\pm1$ edge states co-propagating at the junction. 
 
Above $2\;T$, the QH plateaus are robust (SI-4); interestingly, the nonzero tunneling regime around zero bias appears and becomes increasingly robust as $B$ increases, as is evident from the almost monotonic increase of $2\Delta$ with increasing magnetic field shown in Fig.~\ref{fig4}b. This indicates that as $B$ increases, the spin-polarization of the $\nu_{p,n}=\pm1$ edge-states becomes increasingly non-orthogonal. Note that the 8T responses in Fig.~\ref{fig3}a and ~\ref{fig4}a were taken at different ($V_{BG1}$, $V_{BG2}$) points on the ($-1,1$) plateau (see Figs.~\ref{fig2}b or ~\ref{fig2}(c) and SI-5).

\begin{figure}[ht]
\includegraphics[width=0.48\textwidth]{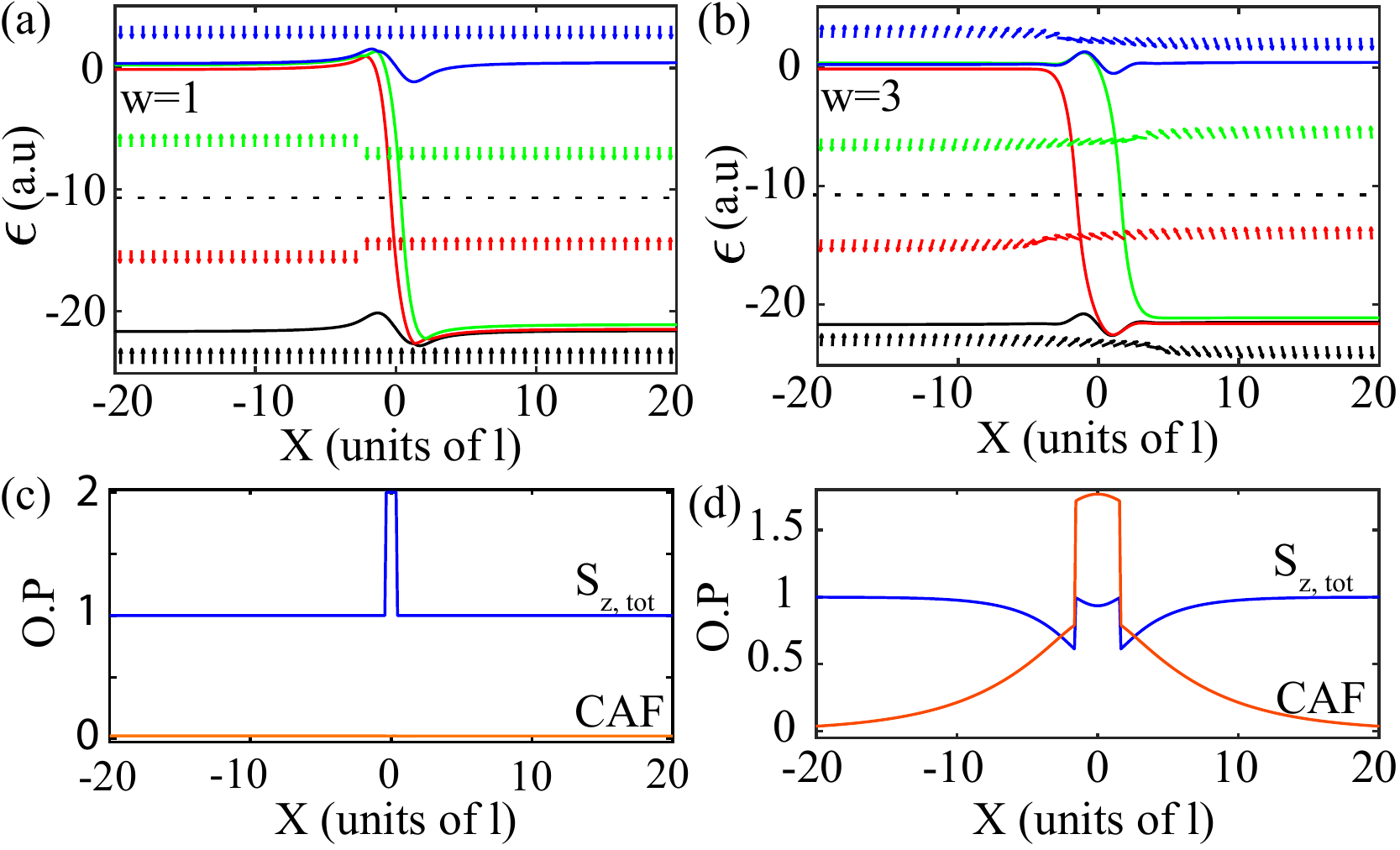}
\caption{\label{fig5}\textbf{Hartree-Fock results:} The parameters used are $E_C=10$ (Coulomb interaction strength), $q_0=0.2$ (screening wavevector), $v_z=0.25$, $v_{xy}=-0.25$ (residual interactions). The interactions are finite-range, with details in the supplemental material. \textbf{(a)} and \textbf{(b)} One-body energies (black, red, green and blue solid lines) for a narrow ($w=\ell$)  and wide ($w=3\ell$) pn junction, respectively, as a function of position $X$ (in units $\ell$). The dashed black line is the chemical potential $\mu$. The colored arrows show (color-coded) spin polarizations  of the one-body states. \textbf{(c)} and \textbf{(d)} show order parameters as a function of position for the narrow and wide junctions, respectively. The narrow junction does not support CAF order, and the spin polarizations of the red and green states at $\mu$ are opposite. The wide junction does support CAF order, with the spin polarizations of the red and green one-body states being non-orthogonal.}
\end{figure}

Our data indicate that at low bias and above a critical field-strength, the two co-propagating edges of the $\nu=0$ strip have non-orthogonal spins. This in turn strongly indicates the presence of CAF order  in the strip. To substantiate this picture, we have carried out Hartree-Fock (HF) calculations. We use a screened Coulomb interaction and finite-range residual interactions~\cite{Das_Kaul_Murthy_2022}, which turn out to be crucial in selecting the ground state in the absence of sublattice anisotropy.  We model the interface by a background charge density rising linearly from $\rho_b(\nu=-1)=1/2\pi\ell^2$ to $\rho_b(\nu=1)=3/2\pi\ell^2$ over a distance $2w$. Details are given in SI-7. The HF results are shown in Fig. \ref{fig5}. The two panels on the left show the results for a narrow junction region of width $w=\ell$, while the two panels on the right are for a wider junction with $w=3\ell$. The one-body HF spectrum  as a function of position for $w=\ell$ is shown in Fig. \ref{fig5}a. Position is measured in units of $\ell$, with the junction being at $X=0$. The far left is the $\nu=-1$ region, where only one of the four ZLLs is occupied, while the far right is the $\nu=1$ region with three of four ZLLs occupied. Two of the states, colored red and green, cross the chemical potential $\mu$ (black dashed line) near the junction. Their spin polarizations are represented by the arrows colored red and green respectively. It is evident that the spin polarizations of the red and green states at the locations where they cross $\mu$ are opposite. Thus, static disorder cannot induce tunneling across the junction, consistent with our data.  Fig. \ref{fig5}c shows the order parameters. The CAF order parameter is identically zero, and the junction is fully spin-polarized. Fig. \ref{fig5}b shows the one-body energy levels for a $w=3\ell$. Once again, the red and green states cross $\mu$, but now their spins (represented by the colored arrows) rotate continuously with position. At the locations where the red/green states cross $\mu$, their spins are almost parallel. Thus, static disorder is able to induce tunneling across the junction. Fig. \ref{fig5}d shows that the CAF order parameter is well-established in the junction, and penetrates several $\ell$ into the bulk on both sides.  

The HF calculations allow us to understand the esperimental data in terms of the dimensionless width of the junction ${\tilde w}=w/\ell$. At small $B$, ${\tilde w}$ is small, and CAF order does not develop, leading to the absence of tunneling. At large $B$ and zero bias, CAF order develops, the two edges have non-orthogonal spins, and disorder-induced tunneling occurs. The $w$ for our device, which is determined by the thickness of the dielectric ($hBN$), is of the order of $\sim$ 30-50 nm and qualitatively agrees with the model. Now with increasing bias we expect that 
bias will lead to charge accumulation across the junction,   causing it to become narrower. Beyond a critical bias voltage $V_b^*$, the junction is expected to collapse to the small $w/\ell$ state, thereby suppressing disorder-induced tunneling across the junction. Also, Fig. \ref{fig4}b shows, $V_b^*$ increases with $B$, i.e. with $w/\ell$, as expected. The precise value of $V_b^*$ depends on many details not included in our present model. The smaller peaks (Fig. 3a) at intermediate bias voltages are not captured in our model. Further studies are required for their origin.


In conclusion, our tunneling data through a $\nu=0$ region sandwiched by $\nu_{p,n}=\pm1$ quantum Hall regions have two broad implications: First, they allow us to infer the presence of CAF order in the $\nu=0$ region when its dimensionless width is sufficiently large. 
Second, the tunneling across the junction  can be controlled by a bias voltage, and in particular can be switched off beyond a critical bias. This implies that the (essentially magnetic) CAF order in the junction is electrically controllable.  Our observations suggest a new way to manipulate the CAF order in $\nu=0$ graphene {\it in situ} for technological purposes, potentially allowing the control of   ultra-fast spin dynamics, coherent transport, and spin superfluidity~\cite{takei2014,takei2016spin}. 

A.D. thanks the Department of Science and Technology (DST), India for financial support (DSTO-2051) and acknowledges the Swarnajayanti Fellowship of the DST/SJF/PSA-03/2018-19. A.D. further thanks for the Institute funding under the Institute of Eminence (IoE). A.D. also thanks Ministry of human resource development (MHRD), India for the financial support under Scheme for Promotion of Academic and Research Collaboration (SPARC/2018-2019/P1178/SL). G.M. is grateful to the US-Israel BSF for partial support under grant no. 2016130, and for the hospitality of ICTS Bangalore during the June 2019 workshop "Edge Dynamics in Topological Phases", where this work was initiated. J.K.J. thanks the U. S. Department of Energy, Office of Basic Energy Sciences for partial support under Grant No. DE-SC-0005042.

$^{\star}$ $^{\dagger}$corresponding authors: 
murthy@g.uky.edu, anindya@iisc.ac.in

%

\pagebreak



\section*{Supplementary material}

# Supplemental Material: Electrically switchable tunneling across a graphene pn junction: evidence for canted antiferromagnetic phase at $\nu = 0$ state

Arup Kumar Paul[1], Manas Ranjan Sahu[1], Kenji Watanabe[2], Takashi Taniguchi[2], J. K. Jain[3], Ganpathy Murthy[4,*], and Anindya Das[1,†]

[1]Department of Physics,Indian Institute of Science, Bangalore, 560012, India.
[2]National Institute for Materials Science, Namiki 1-1, Ibaraki 305-0044, Japan.
[3]Department of Physics, The Pennsylvania State University, University Park, Pennsylvania 16802, USA.
[4]Department of Physics and Astronomy, University of Kentucky, Lexington, Kentucky 40506, USA.

**Supplementary information 1: Device structure**

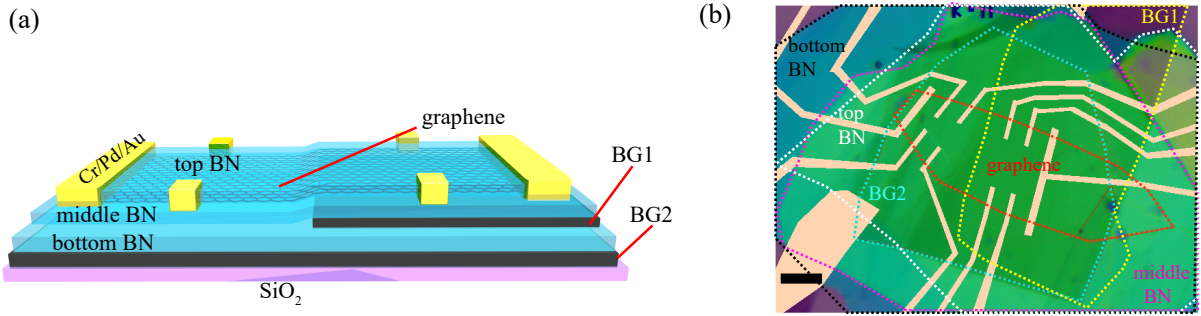

**SI 1:** (a) A schematic of the device. (b) The optical image of the measured device. The flakes are marked with different colors. Scale-bar is $5\mu m$.

Figs. SI 1(a) and SI 1(b) show a schematic of the device structure and an optical image of the device, respectively. In our device, monolayer graphene has a layer of hBN at the top (thickness of $\sim 23nm$), a layer of HBN in the middle (thickness of $\sim 25nm$) that separates the graphene from a graphite back gate BG1, and finally a bottom hBN layer (thickness $\sim 40nm$) that separates the graphite gates BG1 and BG2. These bottom gates partially cover the two halves of the graphene flake, giving independent control of the carrier density in each part. To fabricate the entire stack of encapsulated graphene with graphite gates, we have used the well-known hot pick-up and transfer technique. The edge contacts on the graphene flake were first defined by e-beam lithography and then etched by reactive ion etching (RIE). Subsequently, $Cr(2nm)$, $Pd(10nm)$ and $Au(70nm)$ were deposited on the etched regions by thermal evaporation.

*murthy@g.uky.edu
†anindya@iisc.ac.in

## Supplementary information 2: Measurement setup

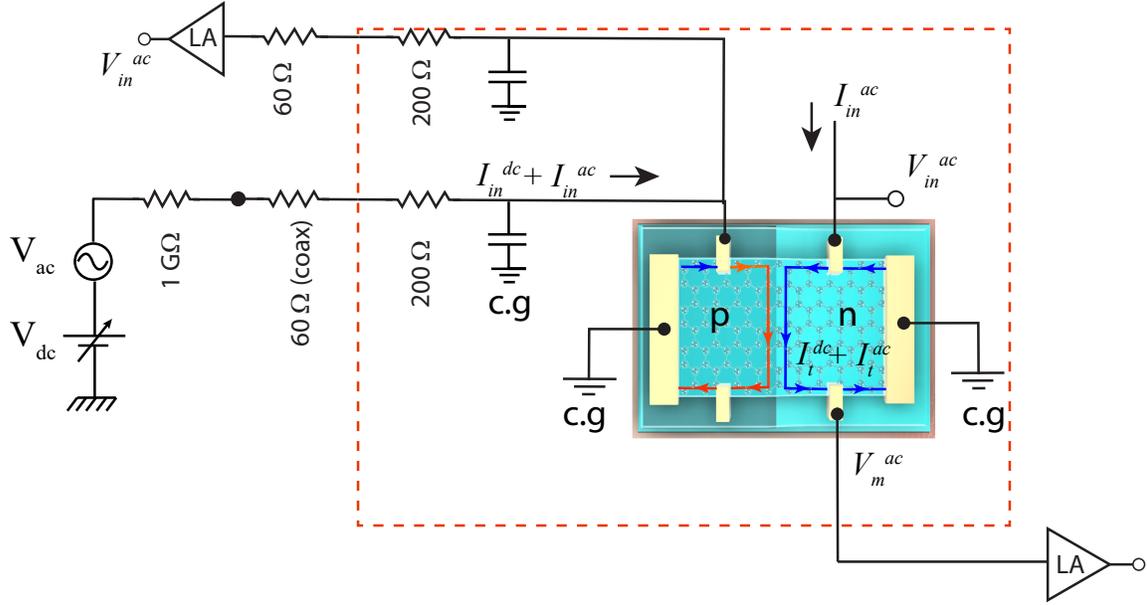

**SI 2:** Complete transmittance measurement setup.

The complete measurement setup is shown in Fig. SI 2. As shown, the ac and dc voltage sources ($V_{ac}$ at 13 Hz and $V_{dc}$, respectively) connect to the current injection contact of the device via a $1G\Omega$ resistance, followed by a coaxial cable ($\sim 60\Omega$) and then a RC filter (R$\sim 200\Omega$, C$\sim 2nF$). While the RC filter and the coaxial cable are inside the dilution fridge, the $1G\Omega$ resistance is located outside the fridge. The $1G\Omega$ resistance acts as the current ballast for the low frequency junction transmittance measurements. The input voltage ($V_{in}^{ac}$) at the current injection point is measured via another RC filter- coax line connected in a $T$-configuration to the injection line. In a similar way, two RC filter- coax lines are also connected to the n side. These $T$-configurations are used for determining the individual quantum Hall responses of the p and n side shown in Fig. SI-4. The ac component of the transmitted current ($I_t^{ac}$) is determined by measuring the voltage drop ($V_m^{ac}$) between the rightmost grounded contact and bottom right Hall bar contact with standard Lock-in technique.

## Supplementary information 3: Differential transmittance and dc transmittance:

In our measurement setup, the differential transmittance ($t_{diff}$) of the junction is measured as a function of applied dc voltage. To derive the junctions' total dc transmittance ($t$), as plotted in Figure 3(a) and Figure 4(a) of the main manuscript, we first calculate the total transmitted current ($I_t$) from the $t_{diff}$ plots

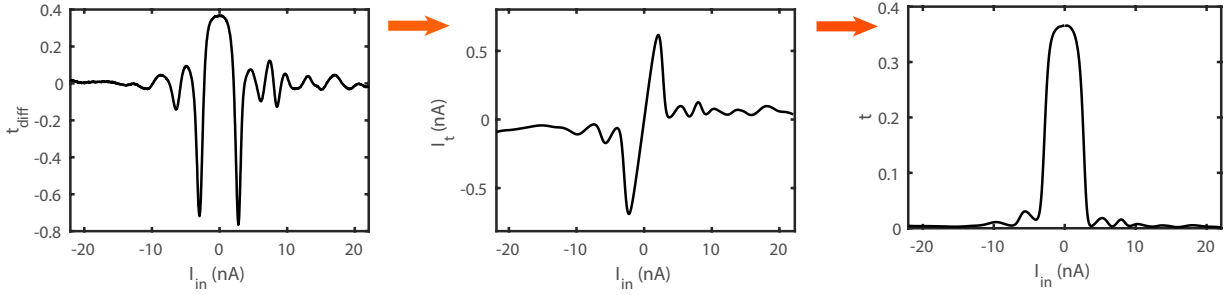

**SI 3:** Transmittance of the junction. Left panel is the measured differential transmittance ($t_{diff}$) as a fucntion of injected dc current ( at 200 mK). Middle panel is the transmitted dc current ($I_t$) through the junction. Right panel is the dc transmission ($t$) of the junction.

as described bellow:

$$\text{Since,} \quad t_{diff} = \frac{I_t^{ac}}{I_{in}^{ac}} = \frac{dI_t}{dI_{in}}$$

$$\implies dI_t = t_{diff} dI_{in}$$

$$\text{therefore,} \quad I_t = \int_0^{I_{in}} t_{diff} dI_{in} \quad (1)$$

Then, the $t$ is derived by simply dividing $I_t$ by total injected dc current $I_{in}$. The Fig. SI 3 shows the calculated $I_t$ and corresponding $t$ with dc bias current at $200 mK$. In the figure, the left-most panel show the measured differential transmittance $t_{diff}$ as function of injected dc current $I_{in}$. The middle panel show the corresponding calculated $I_t$ versus $I_{in}$ plot using Eqn. 1. The right most panel show the $t$ versus $I_{in}$ plot.

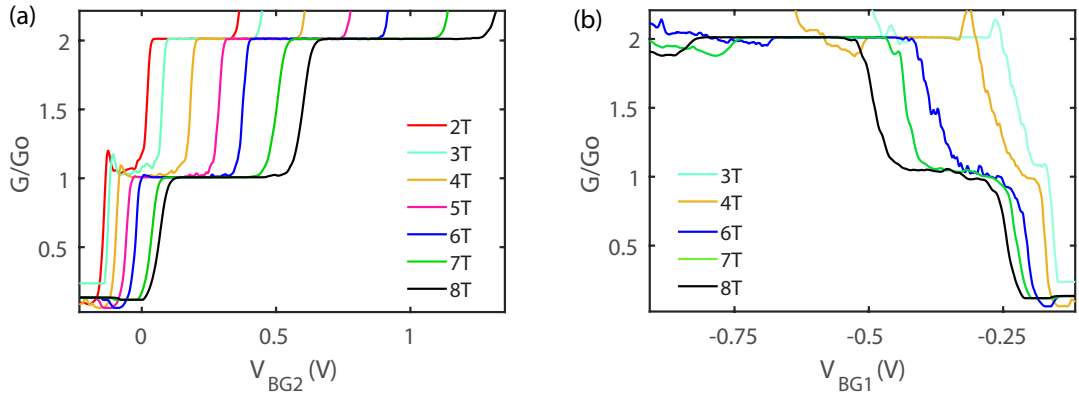

**SI 4:** (a) and (b) quantum Hall responses of the n side and p side, respectively, at 40 mK.

**Supplementary information 4: Quantum Hall response of the device:**

The individual quantum Hall responses of n and p side for different magnetic fields are shown in Fig. SI 4(a) and SI 4(b), respectively. For this measurement, we have kept one side in the insulating regime ($\nu = 0$)

and measure the response of the other side, using the measurement setup described in SI 2. In the figures, conductance ($G$) in units of conductance quanta ($G_0 = e^2/h$) are plotted as function of gate voltages $V_{BG1}$ (p side) and $V_{BG2}$ (n side). For both the cases the broken symmetry integer QH plateaus are visible even at low magnetic fileds.

**Supplementary information 5: Bias dependent tunneling for different Filling factors:**

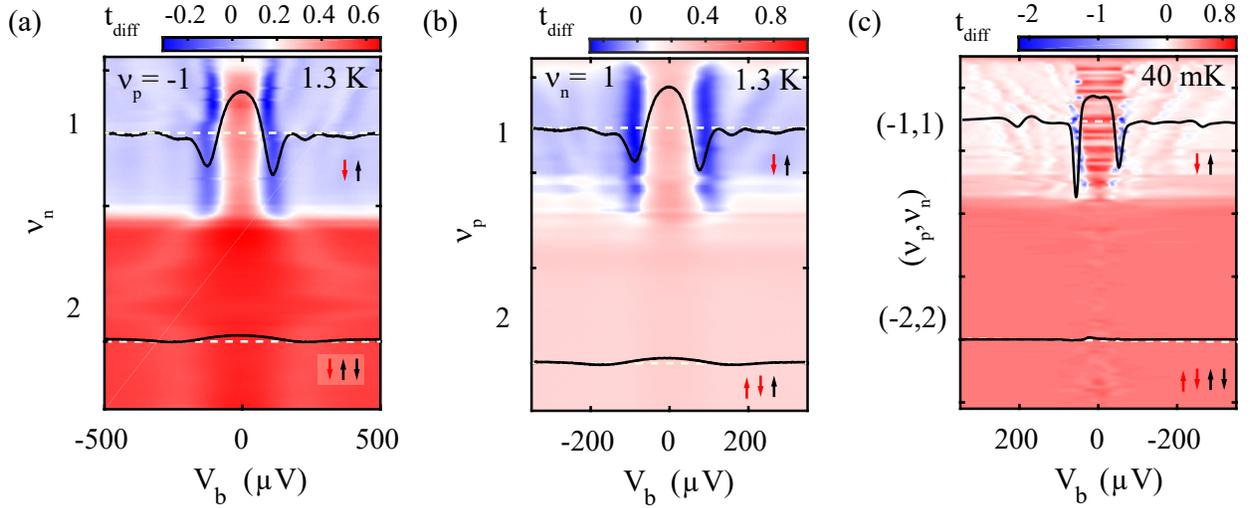

**SI 5:** Filling factor dependence of the bias response. $t_{diff}$ vs. $V_b$ response as functions of (a) n-side filling factor $\nu_n$ and (b) p-side filling factor $\nu_n$ while $\nu_p$=-1 and $\nu_n$=1, respectively. (c) $t_{diff}$ vs. $V_b$ response for $\nu_n = \nu_p$.

The filling factor dependence of the bias response is shown in Fig. SI 5. The Fig. SI 5(a) shows the bias response at different $\nu_n$, while the p-side is kept fixed at $\nu_p = -1$. Similarly, Fig. SI 5(b) shows the bias response at different $\nu_p$, for $\nu_n = 1$. The data was taken at temperature $T = 1.3K$. Fig. SI 5(c) shows the data at $\sim 40mK$, when equal number of edge states co-propagate along the junction, i.e for $|\nu_p| = |\nu_n|$. The insets of the figures show the ideal spin-configuration (↑ and ↓-arrows) of the electron and hole-like edge states (blue and red-arrows, respectively). The bias cut lines along the white dashed lines are also shown in the figures. All the data were taken at magnetic field 8T. As can be seen, the anomalous bias response is only prominent for $(\nu_p, \nu_n) = (-1, 1)$ plateau.

**Supplementary information 6: Full equilibration versus spin selective equilibration:**

Ideally, the equilibration of the symmetry-broken quantum Hall edge states at graphene p-n junction depends on their spin configuration. For example, for $(\nu_p, \nu_n)$ = (-1,1) if the spins are orthogonal, then equilibration

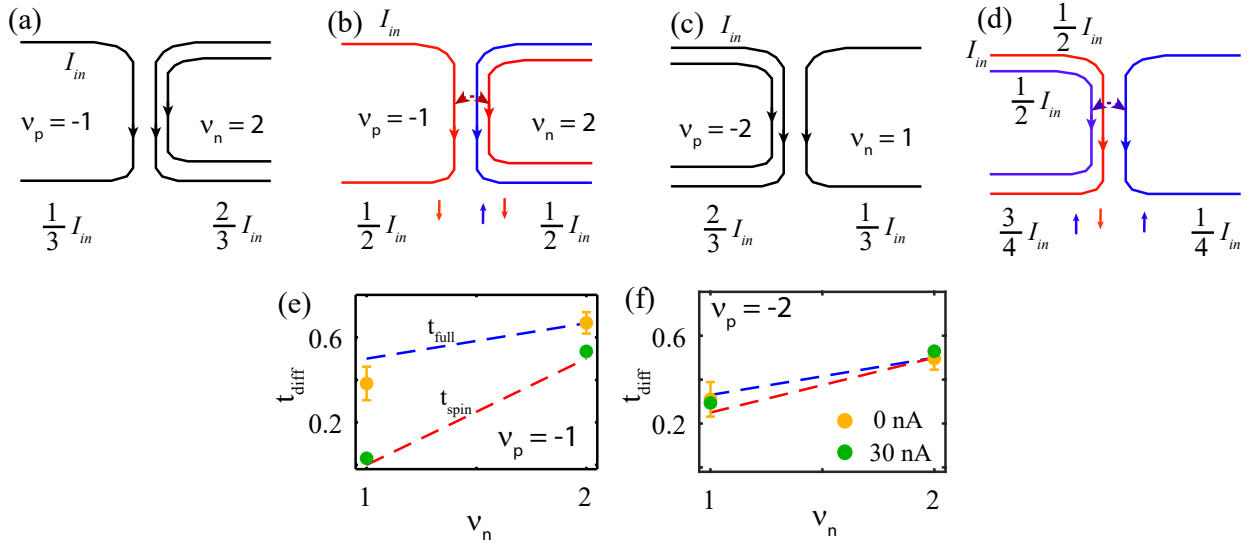

**SI 6:** (a) Full equilibration and (b) spin selective equilibration models for filling factor $(\nu_p, \nu_n)$ = (-1,2). (c) and (d) are same models for filling factor $(\nu_p, \nu_n)$ = (-2,1). (e) and (f) plateau-wise $t_{diff}$ as function of $\nu_n$ for $\nu_p = -1$ and $-2$, respectively. The red and blue dashed lines are theoretically expected transmittance for spin-selective ($t_{spin}$) and full equilibration ($t_{full}$). The green and yellow circles corresponds to the $I_{in}^{dc} = 30$ nA and $I_{in}^{dc} = 0$ nA cases, respectively.

is blocked, as a result transmission of the junction is zero. This is known as spin selective equilibration and can be understood better for higher filling factors having both ↑ and ↓-spin edge states. The Figs. SI 6(a) and SI 6(b), shows the schematic for full and spin selective equilibration, respectively, for filling factor $(\nu_p, \nu_n)$ = (-1,2), where the current $I_{in}$ is injected from the p-side. In case of full equilibration ( Fig. SI 6(a)), the injected current $I_{in}$ is equally distributed among the co-propagating edge states coming from both sides. Hence, after the equilibration the current carried by each of the edge states is $I_{in}/3$. As the $n$ side has $|\nu_n| = 2$ outgoing edge channels, the net transmitted current for full equilibration is $I_t = 2I_{in}/3$. Thus, the net transmittance $t_{full} = 2/3$. For spin selective equilibration (Fig. SI 6(b)), the injected current is only distributed between same spin channels co-propagating along the junction. As a result in this case the transmitted current is $I_t = I_{in}/2$, thus transmittance $t_{spin} = 1/2$. Figs. SI 6(c) and SI 6(d), shows the two cases for filling factor $(\nu_p, \nu_n)$ = (-2,1). Here also, for the full equilibration, the injected current is equally distributed between three channels at the junction, however, as there is only one edge state at the n side, the transmitted current is $I_t = I_{in}/3$, so $t_{full} = 1/3$. For the spin selective case, first the injected current is equally divided between the two up and the single down-spin edge channels at the p side. Then the up and down spin currents gets redistributed spin selectively at the junction. As a result the net transmitted current in this case $I_t = I_{in}/4$, so the transmittance $t_{spin} = 1/4$ [1]. The Figs. SI. 6(e) and SI 6(f) show plateau-wise averaged $t_{diff}$ and its fluctuations (error bars) at 8T and 40 mK, as function of $\nu_n$ keeping $\nu_p$ fixed. In the figures the yellow and green circles corresponds to zero bias ($I_{in}^{dc} = 0$ nA) transmittance

5

value and high bias transmittance value ($I_{in}^{dc} = 30$ nA), respectively. The blue and red dashed lines are the theoretically expected transmittance, for full equilibration ($t_{full}$) and spin-selective equilibration ($t_{spin}$) of the QH edge states [1-3], respectively. It can be seen that for the $(-1, 1)$ and $(-1, 2)$ plateaus, the measured values of $t_{diff}$ for $I_{in}^{dc} = 30$ nA case matches exactly with the spin-selective equilibration picture. However, for the zero-bias case, the measured values are very close to the full equilibration values. For $\nu_p = -2$ case, however, the zero bias and finite bias scenarios are difficult to distinguish.

**Supplementary information 7: Equilibrium Hartree-Fock calculation**

**Model Hamiltonian and bulk ground states for $\nu = 0, \pm 1$:** We will assume that the Hamiltonian has a long-range Coulomb interaction, a short-range density-density interaction, and two other interaction that are Ising-like and $XY$-like in the valley space respectively. In what follows, we quantize the system in Landau gauge, with the spinless single-particle wavefunctions satisfying periodic boundary conditions in the $y$-direction being

$$\Psi_{\alpha,k}(x,y) = \frac{1}{\sqrt{L_y\sqrt{\pi}}} e^{i\alpha K x} e^{iky} e^{-(x-k\ell^2)^2/2\ell^2} \chi_\alpha \tag{2}$$

Here $\alpha = \pm 1 = K, K'$ is the valley index, and $\chi_\alpha$ is a two-component spinor in the sublattice space. In the ZLL, states in the $K$ valley have support only on the $B$ sublattice, and states in the $K'$ valley have support only on the $A$ sublattice. The index $k = 2\pi j/L_y$ ($j$ integer) is the guiding center index, with the wavefunction centered at $k\ell^2$. Given a sample of dimensions $L_x, L_y$ the degeneracy of each Landau level, identical to the number of flux quanta piercing the sample, is $N_\phi = L_x L_y/2\pi \ell^2$. Labelling spin as $s = \uparrow, \downarrow$, our model Hamiltonian in the bulk is

$$\begin{aligned} H &= -E_Z \sum_{\alpha k} c^\dagger_{\alpha\uparrow k} c_{\alpha\uparrow k} - c^\dagger_{\alpha\downarrow k} c_{\alpha\downarrow k} \\ &+ \frac{1}{2L_x L_y} \sum_{\mathbf{q},k,k',s,s',\alpha,\alpha'\beta,\beta'\mu} e^{-iq_x(k-k'-q_y)\ell^2 - (q\ell)^2/2} v_\mu(\mathbf{q}) : c^\dagger_{\alpha s k - q_y}(\tau_\mu)_{\alpha\alpha'} c_{\alpha' s k} c^\dagger_{\beta s' k' + q_y}(\tau_\mu)_{\beta\beta'} c_{\beta' s' k'} : \\ &+ \frac{1}{2L_x L_y} \sum_{\mathbf{q},k,k',s,s',\alpha,\beta} e^{-iq_x(k-k'-q_y)\ell^2 - (q\ell)^2/2} \frac{2\pi e^2}{\kappa \sqrt{q^2 + q_0^2}} c^\dagger_{\alpha s k - q_y} c^\dagger_{\beta s', k' + q_y} c_{\beta s' k'} c_{\alpha,s,k} \end{aligned} \tag{3}$$

where $\mu = 1, 2, 3$ labels Pauli matrices in the valley space. The $U(1)$ symmetry of separate charge conservation in the two valleys implies that $v_1(\mathbf{q}) = v_2(\mathbf{q}) = v_{xy}(\mathbf{q})$. The final term is the long-range Coulomb interaction, assumed to be screened by the small wavevector $q_0$ for computational ease. We keep $q_0$ small $0.2\ell^{-1}$ and verify that nothing qualitative depends on it.

If one assumes that the short-range interactions are ultra-short-range, $v_0, v_{xy}, v_z$ become independent of $\mathbf{q}$. For reasons that will become clear shortly, we will not make this assumption. Instead, we assume the following form in $\mathbf{q}$:

$$v_\mu(\mathbf{q}) = v_{0\mu} e^{-q^2 \xi_\mu^2/2} \tag{4}$$

6

It is important to note that we will allow $\xi_\mu^2$ to be negative, as long as $\xi_\mu^2 > -\ell^2$. As we will see below, negative values of $\xi_\mu^2$ imply that the magnitude of the Fock coupling are greater than that of the Hartree coupling.

We will examine the ground state in the Hartree-Fock (HF) approximation. The most general HF state $|\{\Delta\}\rangle$ is a single Slater determinantal state which can be completely characterized by the expectation values of all 1-body operators

$$\Delta_{\alpha\beta}^{ss'}(k, k') = \langle\{\Delta\}|c_{\alpha sk}^\dagger c_{\beta s'k'}|\{\Delta\}\rangle \tag{5}$$

In the HF approximation, one computes the expectation value of the Hamiltonian in such a state

$$E_{HF}(\{\Delta\}) = \langle\{\Delta\}|H|\{\Delta\}\rangle \tag{6}$$

and looks for the state that minimizes $E_{HF}$. We will restrict our search to states that preserve translation invariance up to a possible intervalley coherence. In other words, we will search among $\Delta$ that satisfy

$$\Delta_{\alpha\beta}^{ss'} = \delta_{kk'}\Delta_{\alpha\beta}^{ss'} \tag{7}$$

Note that the $\Delta$ are independent of guiding center $k$ in the bulk, consistent with translation invariance. By convention, the particle-hole symmetric point is deemed $\nu = 0$, hence the filling of the ZLLs is $f = \nu + 2$. The matrix $\Delta$ is in fact the projector to the set of occupied ZLLs at each guiding center:

$$\Delta = \sum_{j=1}^{f} |\Psi_j\rangle\langle\Psi_j| \tag{8}$$

Each $|\Psi_j\rangle$ is a linear combination of the four available ZLLs, and the set of $|\Psi_j\rangle$ are orthonormal $\langle\Psi_i|\Psi_j\rangle = \delta_{ij}$. The condition on the filling is $Tr(\Delta) = f$.

Given the translation-invariance restriction, and the properties of single Slater determinantal states, we can compute $E_{HF}$ as

$$\begin{aligned}
\frac{E_{HF}}{N_\phi} &= -E_Z \sum_\alpha (\Delta_{\alpha\alpha}^{\uparrow\uparrow} - \Delta_{\alpha\alpha}^{\downarrow\downarrow}) \\
&+ \frac{1}{4\pi\ell^2} \sum_{\alpha\beta ss'} \left(v_{0H} + (-1)^{\alpha+\beta} v_{zH}\right) \Delta_{\alpha\alpha}^{ss} \Delta_{\beta\beta}^{s's'} \\
&- \frac{1}{4\pi\ell^2} \sum_{\alpha\beta ss'} \left(v_{0F} + (-1)^{\alpha+\beta} v_{zF}\right) \Delta_{\alpha\beta}^{ss'} \Delta_{\beta\alpha}^{s's} \\
&+ \frac{1}{\pi\ell^2} \sum_{ss'} \left(v_{xy,H} \Delta_{KK'}^{ss} \Delta_{K'K}^{s's'} - v_{xy,F} \Delta_{KK}^{ss'} \Delta_{K'K'}^{s's}\right) \\
&- E_C \sqrt{\frac{\pi}{2}} \sum_{\alpha\beta ss'} \Delta_{\alpha\beta}^{ss'} \Delta_{\beta\alpha}^{s's}
\end{aligned} \tag{9}$$

7

A number of points should be noted: (i) There is no Hartree term for the Coulomb interaction because of the uniform positive background. $E_C = e^2/\kappa\ell$ is the energy scale of the Coulomb interaction. The properties of projectors ($\Delta^2 = \Delta$) imply that the Coulomb Fock term depends only on the filling, and does not distinguish differently ordered bulk states from each other. (ii) Each short-range interaction has a Hartree coupling $v_{\mu,H}$ and a Fock coupling $v_{\mu,F}$. If the interactions are assumed ultra-short-range, then $v_{\mu,H} = v_{\mu,F}$. However, in our model, given Eq. , we have instead

$$v_{\mu,H} = v_{0\mu}$$
$$v_{\mu,F} = v_{0\mu} \frac{\ell^2}{\xi_\mu^2 + \ell^2} \tag{10}$$

As mentioned earlier, $\xi_\mu^2 < 0$ implies that $|v_{\mu,F}| > |v_{\mu,H}|$. In what follows, it will be convenient to define

$$g_{\mu,H/F} = \frac{v_{\mu,H/F}}{2\pi\ell^2} \tag{11}$$

**CAF ground state at $\nu = 0$:** Much effort has been devoted in the literature to finding the possible ground states at $\nu = 0$, when two linear combinations of the four available ZLLs are filled. Most of the literature assumes ultra-short-range interactions, in which case, as mentioned in the main text, there are four possible phases when no sublattice-symmetry breaking field is present: canted antiferromagnetic (CAF), bond-ordered (BO, sometimes also called Kekule distorted or KD), charge density wave (CDW), and fully spin polarized (F).

More recently, one of the present authors in collaboration with others, has extended the HF treatment by assuming independent Hartree and Fock couplings $v_{\mu,H/F}$. In addition to the phases mentioned above, this more general treatment leads to the coexistence of BO and CAF order parameters when certain conditions are met, namely $g_{xy,F} < g_{xy,H} < 0$, $E_Z > 0$. For experimental samples, it is probably safe to assume $g_{z,H/F} > 0$. As $E_Z$ increases, the coexistent phase undergoes a second-order phase transition to the pure CAF phase, which, at even larger $E_Z$ undergoes another second-order phase transition to the fully polarized phase.

Combining transport and magnon scattering experiments, one can plausibly conclude that CAF order is present in bulk $\nu = 0$ at purely perpendicular field. Therefore, in what follows, we will choose coupling constants consistent with this scenario. The pure CAF state has the following two linear combinations occupied:

$$|\Psi_1\rangle = \cos\frac{\theta}{2}|K\uparrow\rangle + \sin\frac{\theta}{2}|K\downarrow\rangle$$
$$|\Psi_2\rangle = \cos\frac{\theta}{2}|K'\uparrow\rangle - \sin\frac{\theta}{2}|K'\downarrow\rangle \tag{12}$$

The two states have the same average $S_z = \cos\theta$ but opposite values of average $S_x = \pm\sin\theta$. Note that we could introduce a phase into the $\sin\theta/2$ terms, which would rotate the spin in the $xy$ plane, but this does

not affect the energy. Ordering the indices as $K\uparrow$, $K\downarrow$, $K'\uparrow$, $K'\downarrow$, we write the $\Delta$ matrix as

$$\Delta = \frac{1}{2}\begin{pmatrix} 1+\cos\theta & \sin\theta & 0 & 0 \\ \sin\theta & 1-\cos\theta & 0 & 0 \\ 0 & 0 & 1+\cos\theta & -\sin\theta \\ 0 & 0 & -\sin\theta & 1-\cos\theta \end{pmatrix} \quad (13)$$

Using Eq. and omitting the Coulomb energy, the HF energy can now be computed to be

$$\frac{E_{HF}}{N_\phi} = 2g_{0,H} - g_{0,F} - g_{z,F} - 2E_Z\cos\theta - 2g_{xy,F}\cos^2\theta \quad (14)$$

The CAF state occurs only when $g_{xy,F} < 0$ ($E_Z$ is always assumed positive by convention). Minimizing with respect to $\theta$ we obtain the optimal canting angle in the bulk

$$\cos\theta^* = \frac{E_Z}{2|g_{xy,F}|} \quad (15)$$

To explain recent scanning tunneling experiments, a more complicated state with coexistent CAF and bond orders was constructed by one of the present authors and co-workers [4]. Its $\Delta$ matrix is

$$\Delta = \frac{1}{4}\begin{pmatrix} 2+\cos\psi_a+\cos\psi_b & -\sin\psi_a-\sin\psi_b & \cos\psi_a-\cos\psi_b & \sin\psi_a-\sin\psi_b \\ -\sin\psi_a-\sin\psi_b & 2-\cos\psi_a-\cos\psi_b & \sin\psi_b-\sin\psi_a & \cos\psi_a-\cos\psi_b \\ \cos\psi_a-\cos\psi_b & \sin\psi_b-\sin\psi_a & 2+\cos\psi_a+\cos\psi_b & \sin\psi_a+\sin\psi_b \\ \sin\psi_a-\sin\psi_b & \cos\psi_a-\cos\psi_b & \sin\psi_a+\sin\psi_b & 2-\cos\psi_a-\cos\psi_b \end{pmatrix} \quad (16)$$

Such a state can be stabilized for $0 > g_{xy,H} > g_{xy,F}$ and some nonzero range of $E_Z$. Details are in Ref. [4].

**Ground state at $\nu = -1$:** Only a single linear combination of ZLLs is occupied. The Zeeman coupling forces the spin to be polarized, hence the only freedom is in the superposition of valleys. We take the linear combination

$$|\Psi(\theta)\rangle = \cos\frac{\theta}{2}|K\uparrow\rangle + \sin\frac{\theta}{2}|K'\uparrow\rangle \quad (17)$$

Note that we could have introduced a phase to one of the terms, but this phase does not enter the energy. Ordering the indices as $K\uparrow$, $K\downarrow$, $K'\uparrow$, $K'\downarrow$, we write the $\Delta$ matrix as

$$\Delta = \frac{1}{2}\begin{pmatrix} 1+\cos\theta & 0 & \sin\theta & 0 \\ 0 & 0 & 0 & 0 \\ \sin\theta & 0 & 1-\cos\theta & 0 \\ 0 & 0 & 0 & 0 \end{pmatrix} \quad (18)$$

9

The HF energy, ignoring the Coulomb contribution, is straightforwardly computed to be

$$\frac{E_{HF}}{N_\phi} = -E_Z + \frac{1}{2}\Big(g_{0,H} - g_{0,F} + g_{z,H} - g_{z,F} + \sin^2\theta(g_{xy,H} - g_{xy,F} - g_{z,H} + g_{z,F})\Big) \quad (19)$$

If one assumes the $v_\mu$ to be ultra short range, all dependence on $\theta$ disappears, and there is a huge degenearacy of ground states. In our model, the ground state is valley polarized if $g_{xy,H} - g_{xy,F} - g_{z,H} + g_{z,F} > 0$ and an equal superposition of valleys if $g_{xy,H} - g_{xy,F} - g_{z,H} + g_{z,F} < 0$. In real samples, there is usually a sublattice anisotropy (alternatively, valley Zeeman $E_V$) due to partial alignment with the $HBN$ substrate. This favors valley polarization for $\nu = -1$.

**Ground state at $\nu = 1$:** The state has to be spin polarized. An appropriate $\Delta$ with arbitrary valley superposition is

$$\Delta = \frac{1}{2}\begin{pmatrix} 2 & 0 & 0 & 0 \\ 0 & 1+\cos\theta & 0 & \sin\theta \\ 0 & 0 & 2 & 0 \\ 0 & \sin\theta & 0 & 1-\cos\theta \end{pmatrix} \quad (20)$$

The HF ground state energy per guiding center of this state, again ignoring the Coulomb energy, is

$$\frac{E_{HF}}{N_\phi} = -E_Z + \frac{1}{2}\Big(g_{0,H} - g_{0,F} + g_{z,H} - 3g_{z,F} - 4g_{xy,F} + \sin^2\theta(g_{z,F} - g_{z,H} + g_{xy,H} - g_{xy,F})\Big) \quad (21)$$

The $\nu = 1$ state is valley polarized/intervalley coherent under the same conditions as the $\nu = -1$ state.

**HF for a $\nu = 0$ strip sandwiched between $\nu = \pm 1$ regions:** We consider two cases. (i) The $\nu = \pm 1$ regions are valley polarized, and (ii) The $\nu = \pm 1$ regions are intervalley coherent. In both cases the overall phenomenology is the same. If the dimensionless width $\tilde{w} = w/\ell$ of the $\nu = 0$ strip is small (or order 1), there is no CAF order anywhere, whereas if $\tilde{w}$ exceeds a critical value of order unity, the central strip acquires a CAF order parameter, which penetrates for several magnetic lengths into the $\nu = \pm 1$ regions.

The setup is the same as in the bulk, which some key differences. Firstly, one needs to introduce the background potential. This is generated from a background positive charge density with the following form

$$\rho_b(x) = \begin{cases} \frac{1}{2\pi\ell^2} & -\infty < x < -w \\ \frac{1}{\pi\ell^2}(1 + \frac{x}{2w}) & -w < x < w \\ \frac{3}{2\pi\ell^2} & w < x < \infty \end{cases} \quad (22)$$

corresponding to a filling of $\nu = -1$ (one of the four ZLLs filled) on the far left and a filling of $\nu = 1$ (three of the four ZLLs filled) on the far right. This background charge interacts with the electron with the (screened) Coulomb interaction of Eq. 3.

Secondly, since the system is no longer translation invariant in the $x$-direction, one needs to allow the HF averages to vary with position. We will maintain translation invariance in the $y$ direction, leading to

$$\langle c^{\dagger}_{\alpha,s,k} c_{\beta,s',k'} \rangle = \delta_{kk'} \Delta^{ss'}_{\alpha\beta}(k) \qquad (23)$$

Note the dependence of $\Delta$ on $k$, which was not present in the translation invariant bulk.

The HF procedure can now only be implemented numerically. We choose an "active" region of width $120\ell$ around the interface. The active region is bordered by two frozen regions in which the state is fixed by the bulk ($\nu = \pm 1$ on the two sides). The frozen regions contribute mainly a Fock potential to the edges of the active regions. The cylinder that imposes periodic boundary conditions in the $y$-direction is chosen to have circumference $L_y = 20\pi\ell$, such that the separation between successive guiding centers is $\delta X = 2\pi\ell^2/L_y = 0.1\ell$. We use an iterative procedure to implement the HF approximation. One chooses a "seed" configuration, which means a set of $\Delta(k)$ for the entire active region. This is used to generate the HF Hamiltonian, which is diagonalized for each $k$. The chemical potential is determined by overall charge neutrality, and states below the chemical potential are filled. This leads to a new $\Delta(k)$, which completes one iteration. When the absolute value of the difference between the old and new $\Delta$, summed over all matrix elements and all $k$, is lower than some criterion (we used $10^{-4}$), we assume the procedure has converged. All order parameters are determined from the final $\Delta(k)$ by summing over occupied states at every $k$.

**Results:** As we saw, depending on the real-space structure of the residual couplings, the $\pm 1$ bulk states can be either valley polarized or intervalley coherent. We present results for both possibilities, which are qualitatively very similar. The overall finding is that for large $\tilde{w} = w/\ell$, CAF order is present in the central $\nu = 0$ strip, and penetrates several magnetic lengths into the $\pm 1$ regions. For small $\tilde{w}$, all single-particle states have $S_z$ as a good quantum number, and no CAF order exists anywhere. The transition between the two states occurs around $\tilde{w} \approx 1$ in both cases.

*Valley polarized $\nu = \pm 1$:* In the following, energies are in arbitrary units, because only the ratios of couplings matter. In real systems the Coulomb energy scale is of the order of 10s of $meV$, while the residual couplings are a few $meV$. All length scales are in units of $\ell$. We used the following parameter values in this calculation: $E_C = 20$, $q_0 = 0.2$, $g_{z,H} = 0.25$ $\xi_z^2 = -0.2$, $g_{xy,H} = -0.2$, $\xi_{xy}^2 = -0.3$. The negative values of $\xi_z^2$, $\xi_{xy}^2$ ensure that the bulk $\nu = \pm 1$ states are valley polarized even in the absence of sublattice anisotropy.

For $\tilde{w} = 3$, we find the interface to have CAF order in the strip, which penetrates several tens of magnetic lengths into the $\nu = \pm 1$ region. Figure SI 7 shows the HF spectrum, the total order parameters, and the separate order parameters for the four states.

Next we turn to small $\tilde{w}$. Figure SI 8 shows the one-body spectrum for the case $\tilde{w} = 1$ when the strip

11

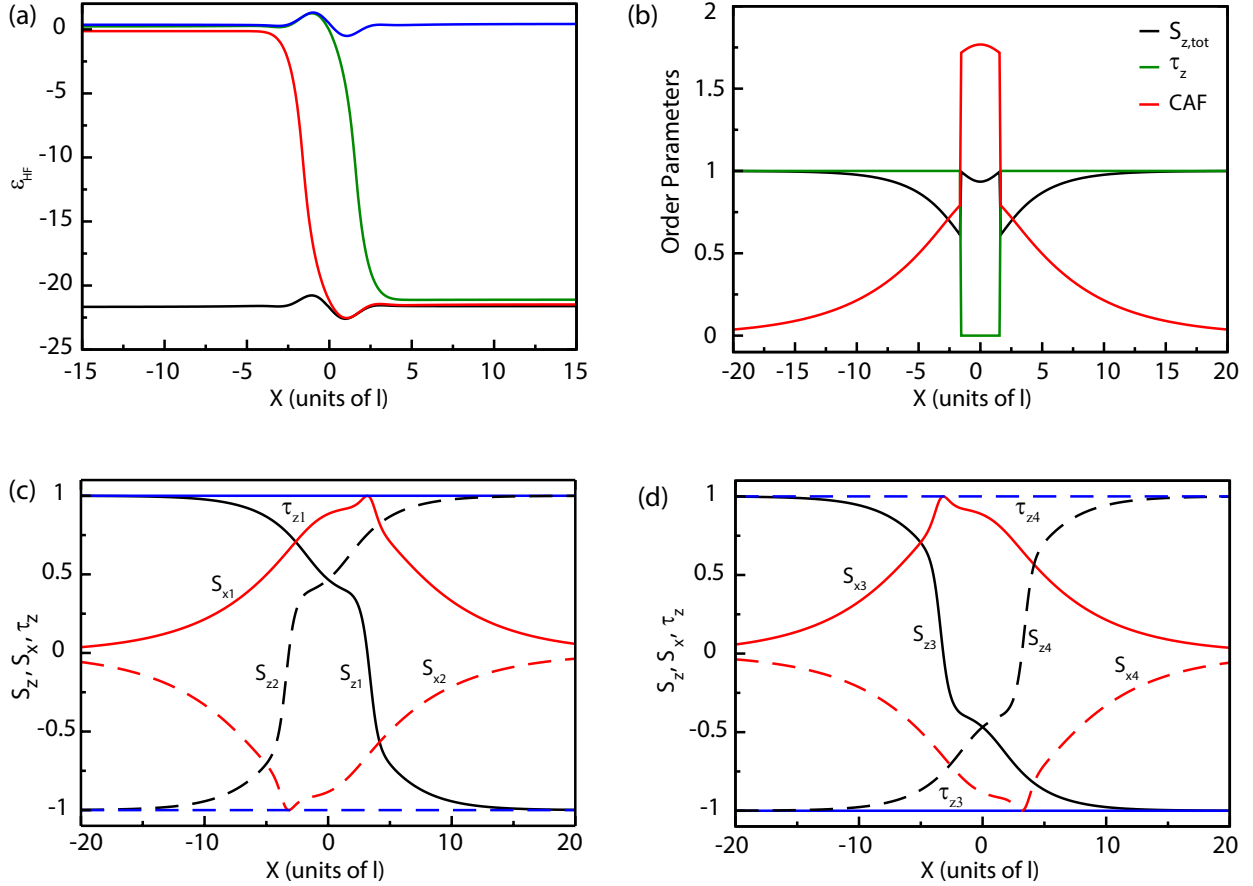

**SI 7:** (a) HF energies for the valley polarized case with $\tilde{w} = 3$. (b) The order parameters vs position. Note the discontinuities near $X = \pm 1.5$ where states cross from above to below the chemical potential. The CAF order parameter is strongest in the region $-1.5 < X < 1.5$, but penetrates quite at least 10 magnetic lengths into the bulk of $\nu = \pm 1$. (c) The order parameters $S_z, S_x, \tau_z$ for the two lower states vs $X$. Solid lines denote the state labelled by the black line in panel (a), while dashed lines denote the state labelled by the red line in (a). The states start out deep in the $\nu = -1$ region as $\uparrow, K$ and $\downarrow, K'$ respectively, but their spin directions rotate continuously in the $xz$ plane. Deep in the $\nu = 1$ region they have exchanged spins, and are now $\downarrow, K$ and $\uparrow, K'$ respectively. (d) Order parameters for the two higher states. Solid lines denote the state labelled by the green line in (a), while dashed lines denote the state labelled by the blue line in (a). Deep in the $\nu = 1$ region, the black, red, and green states of (a) are occupied, and the state is spin and valley polarized.

is in a triplet state. This is evident from SI 8b, which shows the total $S_z$ jumping to 2 in the strip and no CAF order anywhere. In this case states 1 and 2 have $\uparrow$ spin, and states 3 and 4 have $\downarrow$ spin. Thus, there will be no disorder-induced tunneling at low bias for the narrow strip.

12

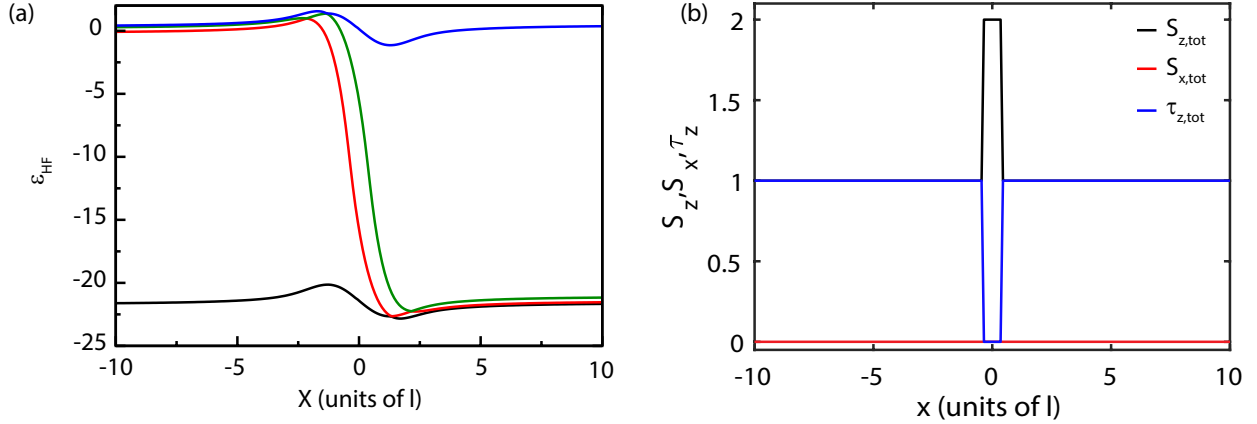

**SI 8:** (a) HF spectrum for the valley polarized $\pm 1$ for small width $\tilde{w} = 1$. (b) Total order parameters. It is clear that there is no CAF order anywhere. The central strip is fully polarized, and thus necessarily valley unpolarized. The state labelled by the red line in (a) is $\uparrow, K'$, while the state labelled by the green line in (a) is $\downarrow, K$. The two edge modes have precisely opposite spin.

*Intervalley Coherent $\nu = \pm 1$:* For completeness we show our results for the case when the couplings are such that $\nu = \pm 1$ are intervalley coherent rather than valley polarized. Here the parameters we use are $E_c = 20$, $q_0 = 0.2$, $g_{z,H} = 2.0$, $\xi_z^2 = 0.1$, $g_{xy,H} = -0.6$, $\xi_{xy}^2 = 0.1$. The positive values of $\xi^2$ ensure that the $\nu = \pm 1$ states are intervalley coherent. In Fig. SI 9 we show the one-body HF spectrum, the total order parameter, and separate order parameters for all the states at $\tilde{w} = 3$. As in the valley polarized case, the CAF order parameter penetrates several magnetic lengths into the bulk. The one-body states now turn out to be a linear combination of a $K$-valley state canted in the $+x, +z$ plane, and a $K'$-valley state canted in the $-x, +z$ plane. Thus, they have no total $S_x$, but each state does have a nonzero average for the CAF order parameter. In Fig. 9(c), we show the $S_z$ and CAF order parameters for states 1 and 2, and in Fig. 9(d) we show the same for states 3 and 4. As before, the spins at the two edges of the strip are not precisely opposite to each other, and thus disorder can induce tunneling between them at low bias.

Finally, we turn to $\tilde{w} = 0.5$ for the same parameters as for $\tilde{w} = 3$. Fig. 10 shows the one-body HF spectrum, and Fig. 10 shows the total order parameter. We note that the strip is fully polarized, and there is no CAF order parameter anywhere. The left edge of the central strip has $\uparrow$ spin while the right edge has $\downarrow$ spin. Thus, there cannot be any disorder-induced tunneling across the strip at low bias.

**Tunneling conductance for long strips:** Assume that the overlap between the spin states on the two edges of the strip is nonzero, so that disorder can induce tunneling across the strip. For simplicity, we will assume that successive disorder-induced tunneling events are uncorrelated, and that the electron loses phase coherence between successive events. Consider a lump of charge travelling along either edge of the

13

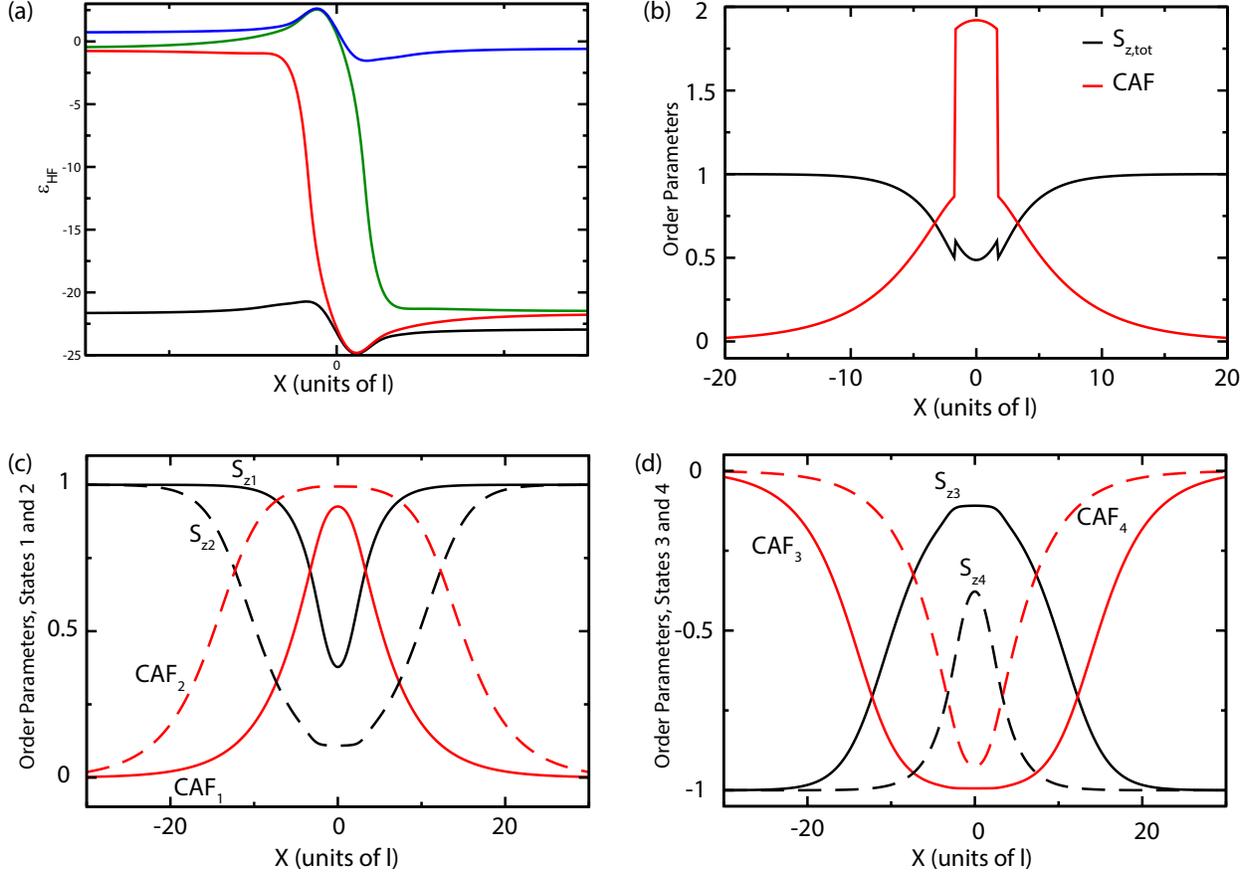

**SI 9:** Intervalley coherent $\nu = \pm 1$ with a $\nu = 0$ strip of large width $\tilde{w} = 3$. (a) HF spectrum. (b) Total order parameters vs $X$. Once again, the CAF order is strongest in the $\nu = 0$ strip, but penetrates at least 10 magnetic lengths into the bulk on either side. Note the discontinuities as the states labelled by the red and green lines in (a) enter the Fermi sea. (c) Order parameters $S_z, S_x\tau_z$ for the two lower states (black and red in (a)) vs $X$. Solid lines are the order parameters of the state labelled by the black line in (a), while dashed lines denote the state labelled by the red line in (a). Each state is a superposition of both valleys in such a way that $\langle S_x \rangle = 0$. However, each state does have a nonzero average $\langle S_x\tau_z \rangle$. (c) Order parameters for the two higher states. Solid lines denote the state labelled green in (a), while dashed lines pertain to the state labelled blue in (a).

central strip. Under the conditions stated above, the charge has a probability $P_l$ to be on the left edge, and a probability $P_r$ to be on the right edge. Let the average tunneling rate be r. Then

$$\frac{dP_l}{dt} = rP_r; \quad \frac{dP_r}{dt} = rP_l \qquad (24)$$

Clearly, $\frac{d}{dt}(P_l - P_r) = -r(P_l = P_r)$. The stable solution is $P_l = P_r = 0.5$. Thus, under the conditions we

14

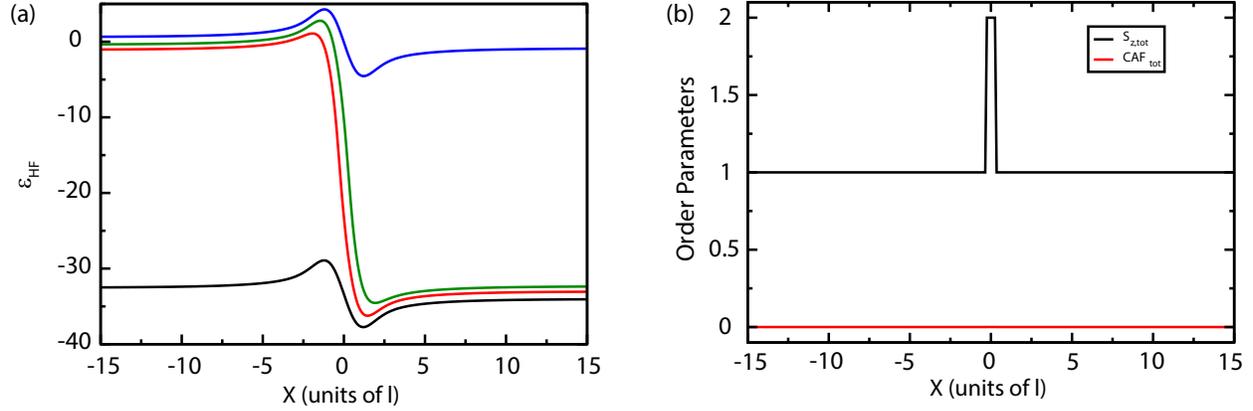

**SI 10:** Intervalley coherent $\nu = \pm 1$ with a $\nu = 0$ strip of small width $\tilde{w} = 1$. (a) HF spectrum. (b) Total order parameters. It is clear that the $\nu = 0$ strip is spin polarized. Thus the red state must be ↑-spin, while the green state must be ↓-spin.

have assumed, half the incident current ends up on each edge.

**Supplementary References:**




\end{document}